\documentclass[conference,10pt]{IEEEtran}

\IEEEoverridecommandlockouts

\usepackage[utf8]{inputenc} 

\usepackage[T1]{fontenc}
\usepackage{url}
\usepackage{ifthen}
\usepackage{cite}
\usepackage{overpic}
\usepackage{balance}
\usepackage[cmex10]{amsmath} % Use the [cmex10] option to ensure complicance
\usepackage{textcase}
\usepackage[tablename=Table]{caption}
\usepackage{blindtext}
\usepackage{floatrow}
\usepackage{soul}
\usepackage{changes}
\usepackage{gensymb}
\usepackage{cite}
\usepackage{amsmath,amssymb,amsfonts}
\usepackage{textcomp}
\usepackage{graphicx}
\usepackage{amssymb}
\usepackage{amsfonts}
\usepackage{amsmath}
\usepackage{epsfig}
\usepackage{epigraph}
\usepackage{color}
\usepackage{fancybox}
\usepackage{textcomp}
\usepackage{multirow}
\usepackage{setspace}
\usepackage{psfrag}
\usepackage{booktabs}
\usepackage{float}
\usepackage[caption = false]{subfig}
\usepackage{algorithm}
\usepackage{algpseudocode}
\usepackage{siunitx}
\usepackage{mathtools, nccmath, bigints, amsfonts}
\usepackage{array}
\usepackage{stfloats}
\newcolumntype{L}{>{\centering\arraybackslash}m{3cm}}

\usepackage{amsthm}
\theoremstyle{definition}

\usepackage{mathrsfs}
\usepackage{mdframed}
\usepackage{tcolorbox}

\newfloatcommand{capbtabbox}{table}[][0.4\textwidth]
\definecolor{matlabgreen}{rgb}{0.0, 0.5, 0.0}

\usepackage{pict2e} 

\theoremstyle{plain}

\newtheorem{lemma}{Lemma}

\newtheorem{corollary}{Corollary}

\usepackage{algpseudocode} % For algorithmic pseudocode commands

\newcommand{\vect}[1]{\mathbf{#1}}

\def\Htran{\mbox{\tiny $\mathrm{H}$}}
\def\Ttran{\mbox{\tiny $\mathrm{T}$}}
\def\CN{\mathcal{N}_{\mathbb{C}}} %Complex Gaussian
\def\imagunit{\mathsf{j}} % Imaginary number

\def\Ncl{N_{\textrm{path}}}

\def\EVM{\mbox{\small $\mathrm{EVM}$}}

\def\imaginary{\mathsf{j}} 
\def\Htran{\mbox{\tiny $\mathrm{H}$}}
\def\Ttran{\mbox{\tiny $\mathrm{T}$}}
    
\begin{document}

\title{Optimizing Movable Antennas in Wideband Multi-User MIMO With Hardware Impairments}

\author{Amna~Irshad\textsuperscript{1}, Emil Bj\"{o}rnson\textsuperscript{1}, Alva~Kosasih\textsuperscript{2}, Vitaly~Petrov\textsuperscript{1}\\
\IEEEauthorblockA{\textit{\textsuperscript{1}KTH Royal Institute of Technology, Stockholm, Sweden} \,\,\, \textit{\textsuperscript{2}Nokia Standards, Espoo, Finland
} \\  Email: \{amnai,emilbjo,vitalyp\}@kth.se, alva.kosasih@nokia.com%
\thanks{The research was supported by the Grant 2022-04222 from the Swedish Research Council and by the SweWIN center (Vinnova grant 2023-00572).}
}

}

\maketitle

\begin{abstract}
Movable antennas represent an emerging field in telecommunication research and a potential approach to achieving higher data rates in multiple-input multiple-output (MIMO) communications when the total number of antennas is limited. Most solutions and analyses to date have been limited to \emph{narrowband} setups. This work complements the prior studies by quantifying the benefit of using movable antennas in \emph{wideband} MIMO communication systems. First, we derive a novel uplink wideband system model that also accounts for distortion from transceiver hardware impairments. We then formulate and solve an optimization task to maximize the average sum rate by adjusting the antenna positions using particle swarm optimization. Finally, the performance with movable antennas is compared with fixed uniform arrays and the derived theoretical upper bound. The numerical study concludes that the data rate improvement from movable antennas over other arrays heavily depends on the level of hardware impairments, the richness of the multi-path environments, and the number of subcarriers. 
The present study provides vital insights into the most suitable use cases for movable antennas in future wideband systems.
\end{abstract}

\begin{IEEEkeywords}
Movable Antennas, Multi-User MIMO, Wideband Channels, Beyond 5G, Error vector magnitude.
\end{IEEEkeywords}

%--------------------------------------------------------------------------------------%
\section{Introduction}
\label{sec:intro}
The Massive MIMO (multiple-input multiple-output) technology has enabled significant improvements in spectral efficiency in 5G networks through efficient spatial multiplexing~\cite{BJORNSON20193}. 
It is built around the so-called \emph{favorable propagation} phenomenon~\cite{Larsson2014a}; that is, the users' channel vectors become statistically nearly orthogonal when the base station (BS) has many more antennas, $M$ than there are active users, $K$. As each additional BS antenna is associated with extra hardware and power costs, it might not be sustainable to maintain the massive antenna surplus (i.e., $M \gg K$) in beyond-5G systems.

To exploit \emph{favorable propagation} and achieve the desired performance, but with a lower number of BS antennas, \emph{intelligent MIMO arrays} with flexible geometry were envisioned in~\cite{BJORNSON20193}. Both hardware technology and algorithms for intelligent MIMO arrays with the so-called \emph{movable antennas} are under active development, as summarized in recent tutorials~\cite{Zhu2025a,ning2024movableantenna}.
As illustrated in~\cite{Wong2021a}, spatial diversity against fading in uplink can be exploited by moving the BS receive antenna to the peak of an impinging waveform at a specific signal frequency. Although this approach leads to notable benefits in narrowband systems, it has limited applicability to modern wideband systems, already equipped with substantial time-frequency diversity~\cite{Lozano2010a}. \emph{Hence, the question of whether or not movable antenna systems can notably improve performance in realistic wideband scenarios remains mostly open.}

Therefore, in this work, we specifically focus on enabling spatial multiplexing of \emph{wideband signals} using a BS equipped with an array of movable antennas, which are few in number but spread over a large aperture. Our goal is to optimize the antenna positions at the BS to achieve a well-conditioned MIMO channel matrix without a massive antenna surplus at the BS (i.e., $M \approx K$) and show when this brings large benefits.

There has been much prior work on sum rate optimization for movable antennas in narrowband systems~\cite{Zhu2025a}. However, these solutions cannot be applied directly to wideband systems because each subcarrier prefers a different solution. Recently, single-user single-antenna wideband movable antenna systems were considered in \cite{hong2025fluidantennaempowering5g} and~\cite{Zhu2024a} using orthogonal frequency-division multiplexing (OFDM). 
However, there is no prior work on wideband MIMO systems with movable antennas.

Hardware impairments in the transceiver limit the practically achievable rates in MIMO systems, particularly at high signal-to-noise ratios (SNR) and when caused by user devices~\cite{massivemimobook}. The detrimental impact of hardware impairments on narrowband downlink multi-user MIMO systems was recently explored in~\cite{yao2024rethinkinghardwareimpairmentsmultiuser}, where it was shown to be the same for fixed and movable antennas. However, the impact of transceiver hardware impairments on \emph{wideband} MIMO system with movable receive antennas remains to be studied.

\subsection{Contributions}

This is the first study of a wideband multi-user MIMO system with movable antennas. The main contributions are:

\begin{itemize}
    \item We derive an \emph{exact OFDM system model} with movable antennas. The impact of the pulse-shaping filter is included, in contrast to the approximate models in \cite{Zhu2024a, Zhu2025a}.
    
    \item We derive the \emph{uplink sum rate} with hardware impairments and study its asymptotic limits as the SNR grows large.

    \item We formulate a \emph{sum rate maximization problem} for the antenna positions and solve it using particle swarm optimization (PSO) methods.

    \item We \emph{identify scenarios where movable antennas provide substantial performance gains}, compared to the conventional uniform planar/linear arrays (UPAs/ULAs), and also when the benefits are negligible. 

\end{itemize}

%--------------------------------------------------------------------------------------%
        \vspace{-5mm}
\section{System Model}
\label{sec:system_model}
\vspace{-0.5mm}

We consider an uplink multi-user MIMO OFDM system with $K$ single-antenna users and a BS with an array of $M$ movable antennas, whose positions are $ \vect{p}_m \in \mathbb{R}^3$ for $m=1,\ldots,M$. The origin of the coordinate system is at the center of the array. We gather these position vectors in the matrix $\vect{P} = [ \vect{p}_1, \ldots,  \vect{p}_M] \in \mathbb{R}^{3 \times M}$, which will be optimized under constraints on how the antennas can be moved.
If a plane wave impinges on the BS from the azimuth angle-of-arrival (AOA) $\varphi$ and elevation AOA $\theta$, the response vector is~\cite[Sec.~7.3.1]{massivemimobook}
\vspace{-0.5mm}
\begin{equation} 
\vect{a}_{\vect{P}}(\varphi,\theta) = \begin{bmatrix} e^{\imaginary \, \vect{p}_1^{\Ttran}\vect{k}(\varphi,\theta) } & \ldots & e^{\imaginary \,\vect{p}_M^{\Ttran}\vect{k}(\varphi,\theta) }
\end{bmatrix}^{\Ttran},
\end{equation}
which depends on the position matrix $\vect{P}$ and the wave vector
\begin{equation} 
\vect{k}(\varphi,\theta) = \frac{2\pi}{\lambda} \begin{bmatrix}\cos(\varphi)  \cos(\theta)  \\  \sin(\varphi) \cos(\theta)\\ \sin(\theta) \end{bmatrix}.
\end{equation}

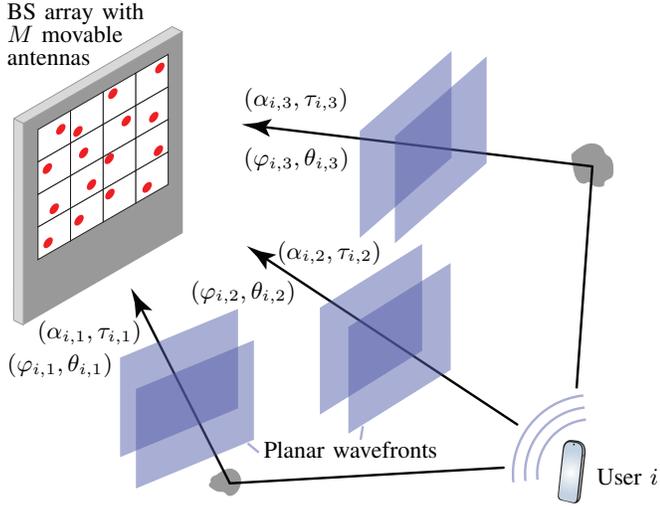
\begin{figure} 
        \centering 
	\begin{overpic}[width=\columnwidth,tics=10]{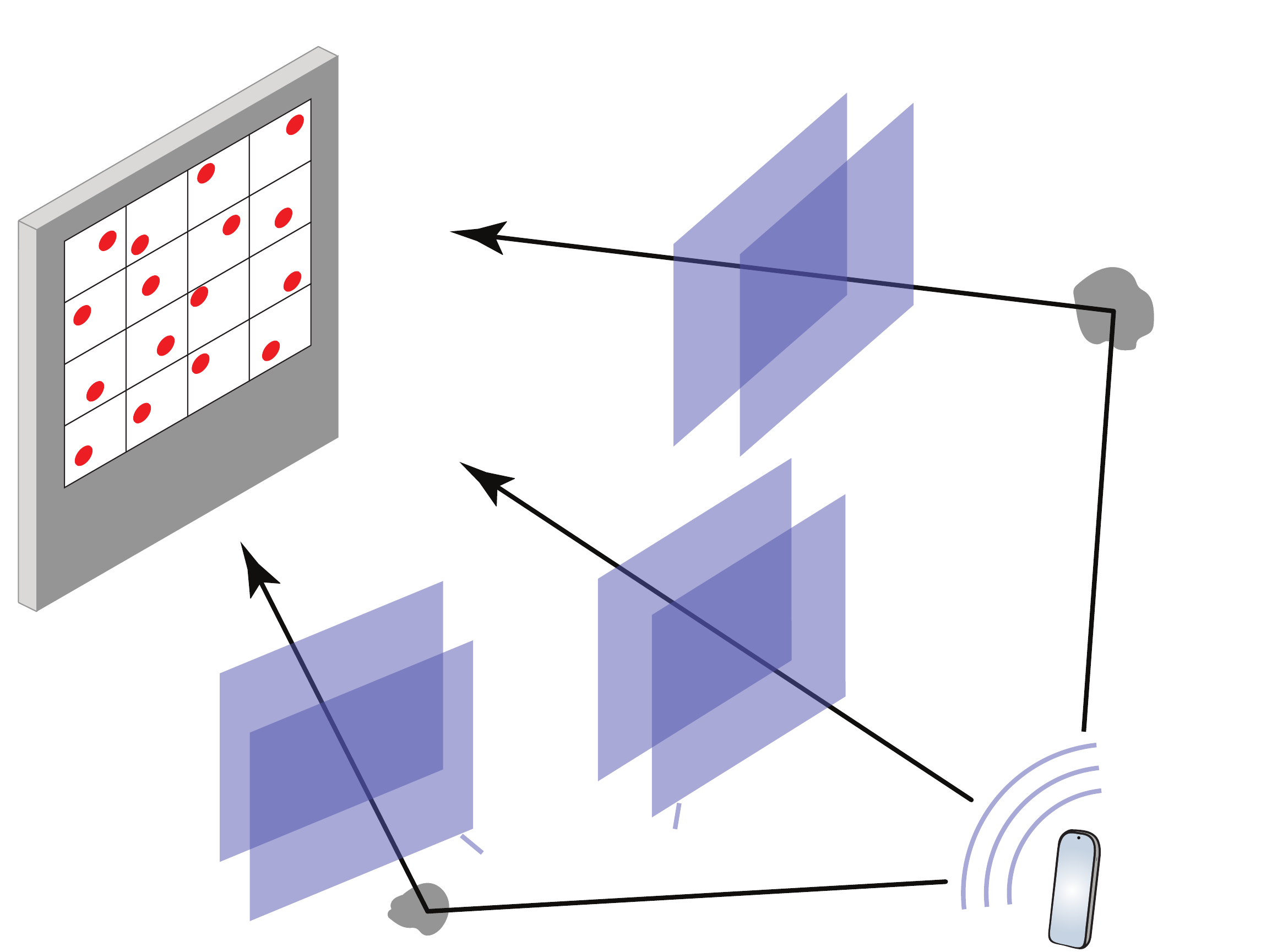}
 \put(0.5,73){\small BS array with}
  \put(0.5,69.5){\small  $M$ movable}
    \put(0.5,66){\small antennas}
    \put(39,7){\small Planar wavefronts}
 \put(89,3){\small User $i$}
   \put(5,25){\small $(\alpha_{i,1},\tau_{i,1})$}
   \put(0.5,20){\small $(\varphi_{i,1},\theta_{i,1})$}
   \put(28,31){\small $(\varphi_{i,2},\theta_{i,2})$}
   \put(41,37){\small $(\alpha_{i,2},\tau_{i,2})$}
   \put(36,51){\small $(\varphi_{i,3},\theta_{i,3})$}
   \put(36,60){\small $(\alpha_{i,3},\tau_{i,3})$}
\end{overpic} 
\vspace{-3mm}
        \caption{Illustration of the considered multipath channel setup.}
\label{figure_MIMO_moveable_multipath} \vspace{-3mm}
\end{figure}

We use a geometric channel model to enable the optimization of $\vect{P}$. The channel from user $i$ has $\Ncl$ far-field paths, where the $n$th path is determined by: 1) amplitude $\alpha_{i,n} \geq 0$; 2) time delay $\tau_{i,n} \geq 0$; 3) azimuth AOA $\varphi_{i,n} \in [-\pi,\pi]$; and 4) elevation AOA $\theta_{i,n} \in [-\pi/2,\pi/2]$. Both the line-of-sight (LOS) path and the scattered paths are modeled like this, as illustrated in Fig.~\ref{figure_MIMO_moveable_multipath}, where the squares represent plane waves.

%%%%%%%%%%%%%%%%%%%%%%
\subsection{OFDM signal model}
The OFDM waveform uses $S$ subcarriers, a subcarrier spacing of $\Delta$, and a pulse-shaping filter $f(t)$ that is only non-zero for $t \in [-1,1]$.\footnote{One example is the triangle function $f(t) = 1-|t|$ for $t \in [-1,1]$ (and $0$ elsewhere), which is obtained by using box functions at the transmitter and receiver. Another example is a time-windowed raised-cosine filter.}
The channel from user $i$ then becomes a finite impulse response (FIR) filter with the taps \cite[Ch.~7]{bookEmil}
\vspace{-0.5mm}
\begin{equation} \label{eq:hk-multitap}
\vect{h}_i[\ell] =  \sum_{n=1}^{\Ncl} b_{i,n}[\ell] \vect{a}_{\vect{P}}(\varphi_{i,n},\theta_{i,n}), \quad \ell=0,\ldots,T,
\end{equation}
where the scalar coefficients $b_{i,n}[\ell] \in \mathbb{C}$ are given by
\vspace{-0.5mm}
\begin{equation}
b_{i,n}[\ell] = \alpha_{i,n} e^{- \imagunit 2 \pi \lambda (\tau_{i,n} - \eta)/c} f \left(\ell+S\Delta(\eta-\tau_{i,n}) \right),
\end{equation}
the speed of light is $c$, the time synchronization coefficient at the receiver is $\eta = \min_{i,n} \tau_{i,n}$  (matched to the fastest path), $\lceil\cdot \rceil$ is the ceiling operation, and the number of delay taps is $T = \left\lceil S\Delta (\max_{i,n} \tau_{i,n} - \eta) \right\rceil$. The number of taps increases with the bandwidth $S\Delta$, so that the paths become more distinguishable.

The received uplink signal $\bar{\vect{y}}[\nu] \in \mathbb{C}^M$ on subcarrier $\nu$ is
\vspace{-0.5mm}
\begin{equation} \label{eq:received-signal-MIMO-OFDM}
\bar{\vect{y}}[\nu] =  \sum_{i=1}^{K} \bar{\vect{h}}_i [\nu] \bar{\chi}_i[\nu] + \bar{\vect{n}}[\nu], \quad \nu =0,\ldots,S-1,
\end{equation}
where $\bar{\chi}_i[\nu]$ is the signal transmitted by user $i$ with power $\rho$, $\bar{\vect{n}}[\nu] \sim \CN ( \vect{0}, \sigma^2 \vect{I}_M)$ is independent noise, and 
\begin{equation} \label{eq:H-frequency-domain-multipath}
\bar{\vect{h}}_i[\nu] = \sum_{n=1}^{\Ncl}  \left( \sum_{\ell = 0}^{T} b_{i,n}[\ell]  e^{-\imaginary 2 \pi \ell \nu /S}\right)
\vect{a}_{\vect{P}}(\varphi_{i,n},\theta_{i,n}), 
\end{equation}
is the channel from user $i$ obtained by taking the DFT of \eqref{eq:hk-multitap}.

%%%%%%%%%%%%%%%%%%%%%%
\vspace{-0.5mm}
\subsection{Communication Performance with Hardware Distortion}
\vspace{-0.5mm}
Practical user devices are affected by various hardware impairments, which in totality can be quantified by the error vector magnitude (EVM) \cite[Sec.~6.1]{massivemimobook}. To capture these effects, we model the transmitted signal by user $i$ on subcarrier $\nu$ as
\begin{equation} \
\bar{\chi}_i[\nu] = \sqrt{\rho} \left(\sqrt{1-\EVM^2} d_i[\nu] + \epsilon_i[\nu]\right),
\end{equation}
where $d_i[\nu]$ is the distortionless unit-variance data signal and $\epsilon_i[\nu]$ is the uncorrelated additive distortion noise with variance $\EVM^2$. The total transmit power is $\mathbb{E} \{ |\bar{\chi}_i[\nu]|^2\} =  \rho ( 1-\EVM^2 +\EVM^2) = \rho$ regardless of the EVM, but the coefficient $\EVM \in [0,1]$ determines the fraction of distortion.

\begin{lemma} \label{lemma:sumrate}
If the receiver knows the channel, an achievable average sum rate over the subcarriers is
\begin{align} \nonumber
R_{\Sigma} &= \frac{1}{S} \sum_{\nu=0}^{S-1} \log_2 \det \left( \vect{I}_M + \frac{\rho}{\sigma^2} \bar{\vect{H}}[\nu] \bar{\vect{H}}^{\Htran}[\nu] \right) \\
& \quad -  \frac{1}{S} \sum_{\nu=0}^{S-1} \log_2 \det \left( \vect{I}_M + \frac{\rho \EVM^2}{\sigma^2} \bar{\vect{H}}[\nu] \bar{\vect{H}}^{\Htran}[\nu] \right), \label{eq:sumrate-expression}
\end{align}
where $\bar{\vect{H}}[\nu]  = [\bar{\vect{h}}_1[\nu] , \ldots, \bar{\vect{h}}_K[\nu] ]$ denotes the channel matrix.
\end{lemma}
\begin{IEEEproof}
The received signal in \eqref{eq:received-signal-MIMO-OFDM} has the same form as a point-to-point MIMO system with the channel matrix $\bar{\vect{H}}[\nu] $, the transmit covariance matrix $\rho (1-\EVM^2) \vect{I}_K$, and the uncorrelated noise term $\sum_{i=1}^{K} \bar{\vect{h}}_i [\nu]  \sqrt{\rho}\epsilon_i[\nu] + \bar{\vect{n}}[\nu]$ with the covariance matrix
$\vect{Q}[\nu] = \rho \EVM^2 \bar{\vect{H}}[\nu] \bar{\vect{H}}^{\Htran}[\nu] + \sigma^2 \vect{I}_M$.
Hence, we can use the worst-case uncorrelated additive noise theorem from \cite{Hassibi2003a} to achieve the following sum rate on subcarrier $\nu$:
\begin{align} \nonumber
R [\nu] &= \log_2 \det \left( \vect{I}_M + \rho (1\!-\!\EVM^2) \vect{Q}^{-1}[\nu] \bar{\vect{H}}[\nu] \bar{\vect{H}}^{\Htran}[\nu] \right) \\ \nonumber
&= \log_2 \det \left( \frac{\rho (1\!-\!\EVM^2)}{\sigma^2}  \bar{\vect{H}}[\nu] \bar{\vect{H}}^{\Htran}[\nu] + \frac{1}{\sigma^2} \vect{Q}[\nu] \right) \\
& \quad - \log_2\det \left( \frac{1}{\sigma^2} \vect{Q}[\nu] \right).
\end{align}
By averaging over the $S$ subcarriers and simplifying the expression, we obtain the average sum rate in \eqref{eq:sumrate-expression}.
\end{IEEEproof}

The sum rate expression in Lemma~\ref{lemma:sumrate} has two terms, where the first represents the sum rate with ideal hardware and the second is the penalty imposed by hardware impairments. 
The EVM determines a hardware-imposed upper bound on the sum rate, which can be seen in the high-SNR regime.

\begin{corollary} \label{cor1}
If $\bar{\vect{H}}[\nu]$ has rank $K$ and $\EVM>0$, then
\vspace{-0.5mm}
\begin{align} \label{eq:upper-limit}
\lim_{\rho \to \infty} R_{\Sigma} =  K \log_2 \left(\frac{1}{\EVM^2} \right).
\end{align}
\end{corollary}
\begin{IEEEproof}
Let $\vect{\Lambda}[\nu]$ be a $K \times K$ diagonal matrix containing the non-zero eigenvalues of $\bar{\vect{H}}[\nu] \bar{\vect{H}}^{\Htran}[\nu]$. We can rewrite \eqref{eq:sumrate-expression} as
\vspace{-0.5mm}
\begin{align} \nonumber
R_{\Sigma} &= \frac{1}{S} \sum_{\nu=0}^{S-1} \log_2 \det \left(  \frac{\sigma^2}{\rho} \vect{I}_K + \vect{\Lambda}[\nu] \right) \\
& \quad -  \frac{1}{S} \sum_{\nu=0}^{S-1} \log_2 \det \left( \frac{\sigma^2}{\rho}\vect{I}_K +  \EVM^2 \vect{\Lambda}[\nu] \right) \\
& \to \frac{1}{S} \sum_{\nu=0}^{S-1} \left(\log_2 \det \left( \vect{\Lambda}[\nu] \right) - \log_2 \det \left( \EVM^2\vect{\Lambda}[\nu] \right) \right)\label{upper}
\end{align}
when $\rho \to \infty$. This limit can be simplified to \eqref{eq:upper-limit}.
\end{IEEEproof}

This upper bound is independent of the BS antenna positions, but their values will determine the performance at finite SNRs; for example, how quickly the bound is approached when $\rho$ increases.
We will optimize the antenna positions next.

%--------------------------------------------------------------------------------------%
\section{Problem Formulation and Solution}
\label{sec:problem-formulation}
Since the BS is equipped with movable antennas, it can optimize the $M$ antenna positions 
$(\vect{p}_1,\ldots,\vect{p}_M)$ to maximize the sum rate $(R_{\Sigma})$ defined in \eqref{eq:sumrate-expression}. We assume antenna $m$ can be moved within a specific region $\mathcal{C}_m \subset \mathbb{R}^3$ as in the prior works \cite{Zhu2024a,hong2025fluidantennaempowering5g} and formulate our optimization problem as follows:
\begin{align} \label{optproblem}
\underset{\vect{p}_1,\ldots,\vect{p}_M}{\text{maximize}} \quad & R_{\Sigma}(\vect{p}_1,...,\vect{p}_M), \\ \label{cona}
\text{subject to} \quad & {\vect{p}_m} \in \mathcal{C}_m, \quad 1\leq m \leq M, \\ \label{conb}
& {\| \vect{p}_m-\vect{p}_j \|}_2 \geq \lambda/2, \quad 1\leq m \neq j \leq M,
\end{align}
where (15) sets antennas $\geq$$\lambda/2$ apart to avoid mutual coupling.

We consider a sparse planar array where each antenna can be moved in non-overlapping 2D squares in the $yz$-plane, as illustrated in Fig.~\ref{figure_MIMO_moveable_multipath}. Each square as size $L \times L$ and the center points are denoted as $(0,y_{m}^0,z_{m}^0)$, for $m=1,\ldots,M$. 
Hence, the set of possible positions for antenna $m$ is
\vspace{-0.5mm}
\begin{equation}
\mathcal{C}_m = \left\{ (0,y, z) \, \Big|\, |y-y_{m}^0|  \leq\frac{L}{2}, \,  | z-z_{m}^0| \leq \frac{L}{2} \right\}.
\end{equation}

\subsection{Proposed Solution Algorithm}
In this section, we present our proposed algorithm for optimizing antenna positions to solve the problem formulated in \eqref{optproblem}. We utilized the PSO method~\cite{PSObookClerc} implemented using the Yarpiz toolbox \cite{kalami2020ypea}, similar to our previous work \cite{PIA2025} that considered narrowband systems. In essence, PSO considers $N_{\textrm{pt}}$ candidate solutions, known as particles, where each particle represents a unique configuration of the antenna positions. The search space is defined by the constraints in \eqref{cona} and \eqref{conb}. 
Each particle updates its position iteratively based on both its individual best-known position and the swarm's global best-known position, following five main steps:

\begin{enumerate}
    \item Initialize the particles randomly but satisfying \eqref{cona}--\eqref{conb}.
    \item Evaluate the sum rate in Lemma~\ref{lemma:sumrate} for each particle.
    \item Update the individual and global best-known positions to keep track of the largest sum rate values.
    \item Update the position and so-called velocity of each particle based on best-known positions and recent movements; see \cite[Eq.~(18)--(19)]{PIA2025} for details.
    \item Terminate the PSO algorithm if the maximum number of iterations is reached, otherwise, return to Step 2.
\end{enumerate}
The sum rate increases monotonically in each iteration and the global best-known antenna positions are selected when the algorithm terminates ($100$ iterations in our numerical study).

%--------------------------------------------------------------------------------------%
\section{Numerical Results}
\label{sec:simulations}
\begin{figure}[!t]
        \centering 
	\begin{overpic}[width=\columnwidth,tics=10]{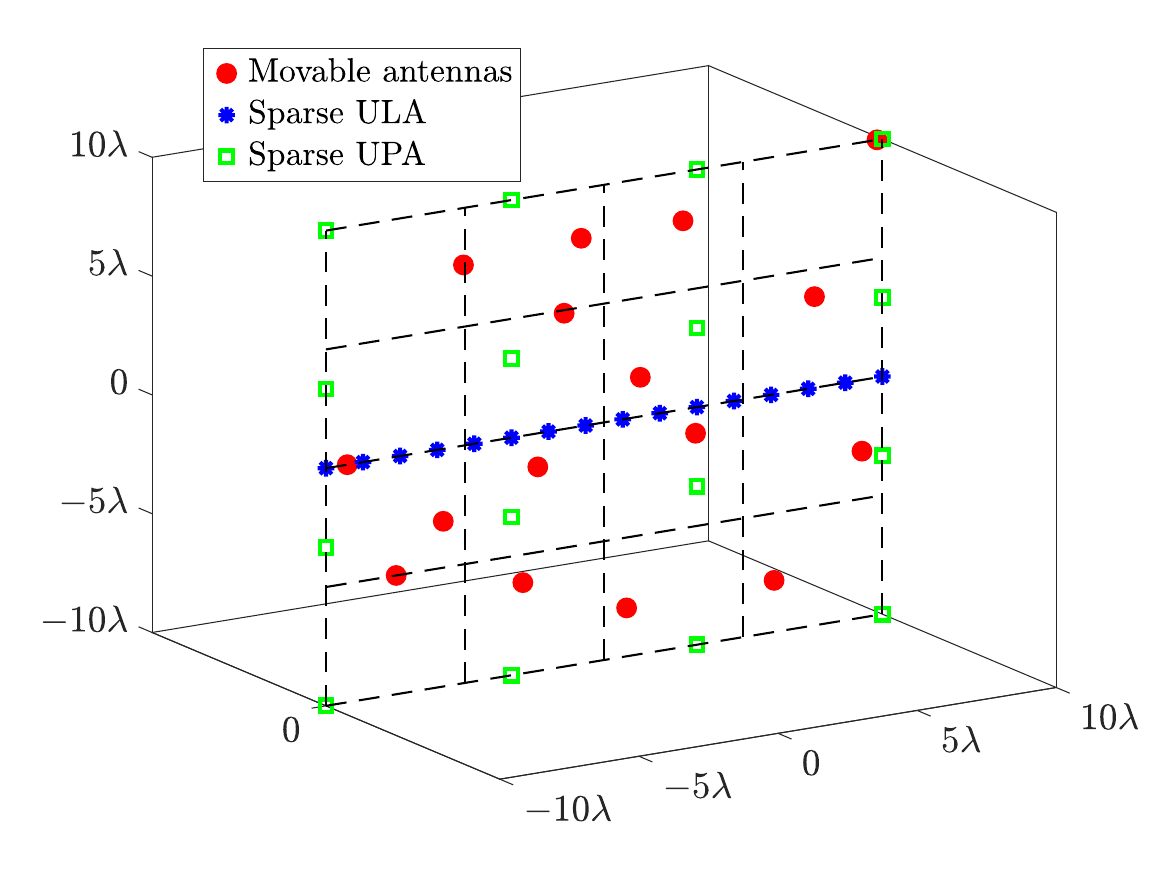}
\end{overpic} \vspace{-8mm}
        \caption{Regions where the movable antennas can be optimized and the antenna positions with a sparse UPA or a sparse ULA.}
\label{figure_arrayExample} \vspace{-2mm}
\end{figure}

\vspace{-1mm}
We evaluate the performance gains of movable antennas compared to conventional fixed antenna arrays of different shapes. We consider a BS array with $M$ movable antennas, where each can move freely within a square of size $5\lambda \times 5\lambda$. The squares are nonoverlapping and positioned on a $4 \times 4$ grid (Fig.~\ref{figure_arrayExample}), where the red dots indicate one ``optimized'' selection of the antenna positions. Key parameters are given in Table~\ref{Parameters}. 

\begin{table}[b!]
    \centering \vspace{-4mm}
    \caption{Summary of simulation parameters}
    \label{Parameters}
   \begin{tabular}{ p{0.55cm}||p{4.4cm}||p{2cm}}
    \hline
    \textbf{Var.} & \textbf{Description} & \textbf{Default}\\
    \hline
    $K$ & Number of users & $10$   \\
    $\sigma^2$& Noise variance (pW) & $3.98$   \\
    $M$ & Number of BS antennas & $16$\\
    $\Delta$&Subcarrier spacing (kHz)& 15\\
    $f_c$ & Carrier frequency (GHz) & 3 \\
    $N_{\textrm{pt}}$ & Number of particles in PSO& 180\\
    $\rho$ & Transmit power (mW/MHz)&1 \\
    $r$ & Radial distance for users (m) & $\mathcal{U}(100,300)$  \\
    $\phi$ & Azimuth angle for users (rad) &  $\mathcal{U}(-\pi/3,\pi/3) $ \\
    \hline
    \end{tabular} \vspace{-3mm}
\end{table}

The BS is mounted on a wall with its center at a height of $4$ meters, defining the origin of the coordinate system. It serves $K=10$ outdoor users, each randomly placed in a region $(r\cos(\phi),r\sin(\phi),-2.75)$, at a height of $1.25$ meters above the ground. 
We will compare the average sum rate achieved by an optimized array (using the methodology of Section~\ref{sec:problem-formulation}) with three baselines with fixed arrays:
\begin{enumerate}
    \item \textbf{Sparse UPA}: It has the maximum antenna spacing achievable within the same aperture as movable antennas and is illustrated with green squares in Fig.~\ref{figure_arrayExample}.
    \item \textbf{Sparse ULA}: This is a ULA where all the antennas are deployed horizontally in the middle of the considered aperture, as shown with blue stars in Fig.~\ref{figure_arrayExample}.
    \item \textbf{Compact UPA}: This UPA is placed around the origin with a classic compact antenna spacing of $\lambda/2$.
\end{enumerate}

%%%%%%%%%%%%%%%%%%%%%
\subsection{Impact of Propagation Environment}
We will first explore how the channel characteristics affect the preferred array geometry in wideband systems with a varying number of subcarriers ($S$) and benchmark the considered arrays against an interference-free upper bound, i.e., optimized movable antennas with no multiuser interference in \eqref{eq:sumrate-expression}.

\begin{figure} 
        \centering 
        {\includegraphics[width=1\textwidth]{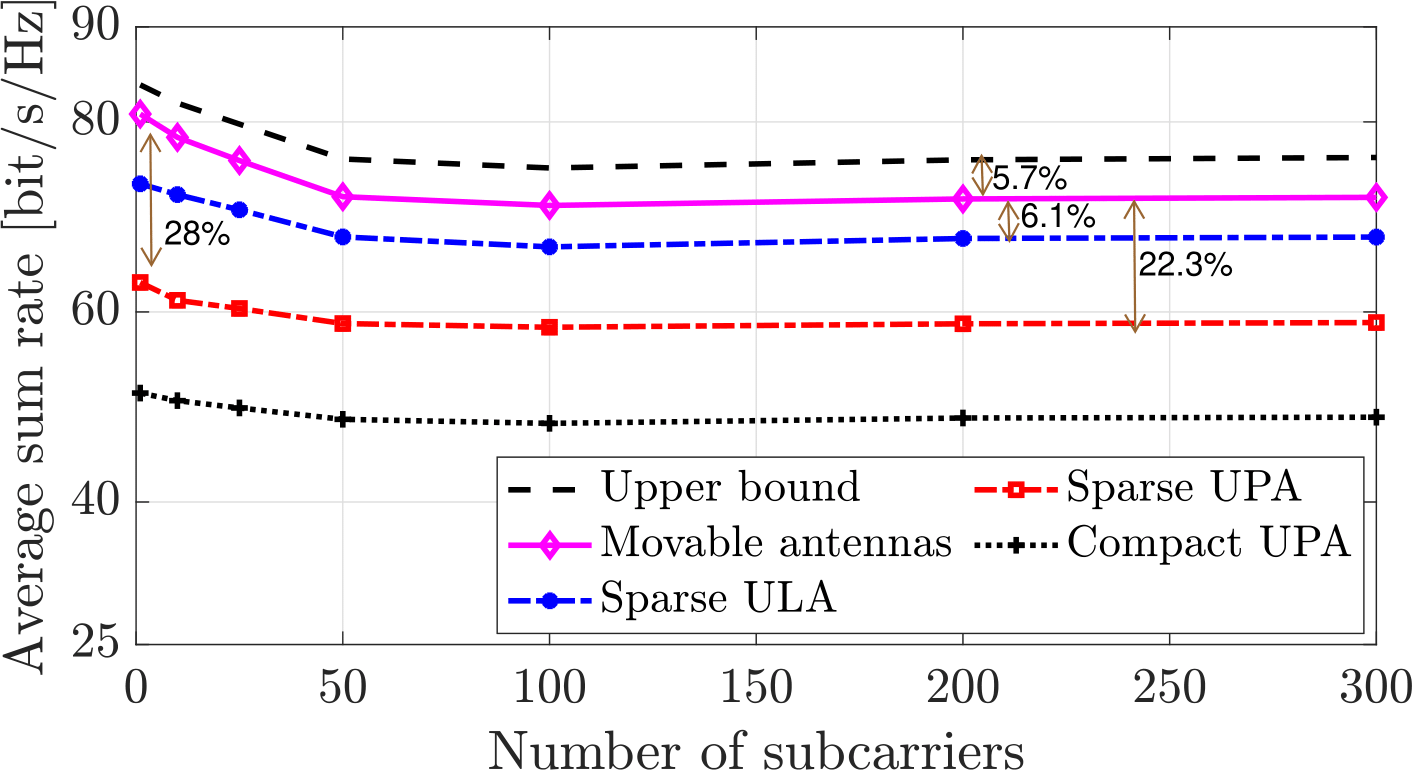}}
        \caption{Average sum rate with $K=10$ users for a varying number of subcarriers, $S$, in a LOS-dominant scenario.}
\label{figure_simulationLOScase} 
\end{figure}

In Fig.~\ref{figure_simulationLOScase}, we consider a LOS-dominant propagation environment inspired by the urban microcell model in \cite[Sec.~5.3.2]{3GPP25996}, which is used to calculate path losses and generate propagation paths in the far field.
There is a LOS path and $6$ scattering clusters, each consisting of $20$ discrete paths.\footnote{The clusters are distributed uniformly in a $\pm 40^\circ$ azimuth interval and $\pm 20^\circ$ elevation interval around the LOS angles. The paths are spread in a $\pm 5^\circ$ interval around the cluster center. The delays are up to $10$ times longer than for the LOS path and their strengths are computed as in \cite[Sec.~5.3.2]{3GPP25996}.} The Rician $\kappa$-factor is $10$\,dB and $\EVM=0.02$. The figure shows the average sum rate for different user realizations for $1$ to $300$ subcarriers.

The figure shows large sum rate variations depending on the array geometry. The highest rates are achieved with optimized movable antennas. The proposed PSO algorithm finds excellent antenna positions, which are only 3$\%$ from the interference-free upper bound. Movable antennas achieve rates up to $57\%$ larger than with a compact UPA and up to $28\%$ larger than with a sparse UPA, which highlights that the gains originate from both increasing the aperture and fine-tuning the antenna positions.
The sparse ULA performs relatively closely to the movable antennas. The reason is that the users are distributed randomly in the azimuth plane and since the propagation is LOS-dominant, all channel components have roughly the same elevation angles. The sparse ULA achieves the maximum spatial resolution in the azimuth plane, so it is well-designed for this scenario even if the antennas are fixed.
Additionally, we notice that increasing the number of subcarriers leads to a slight reduction in the rate per subcarrier. The loss is largest with movable antennas, but the performance differences between the array types are nearly maintained.

\begin{figure} 
        \centering 
        {\includegraphics[width=1\textwidth]{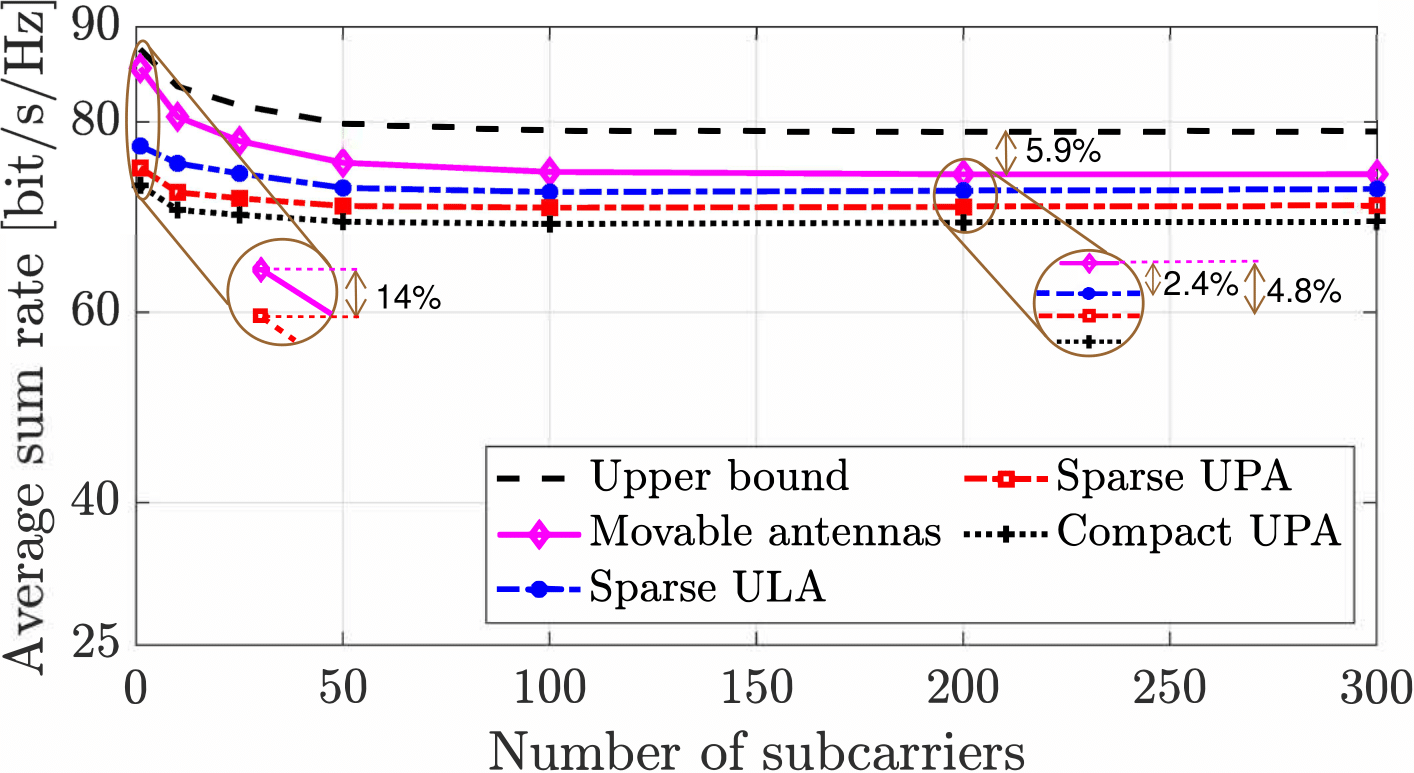}}
        \caption{Average sum rate with $K=10$ users for a varying number of subcarriers, $S$, in a rich non-LOS scenario.}
\label{figure_simulationNLOScase} 
\end{figure}

In Fig.~\ref{figure_simulationNLOScase}, we instead consider a rich scattering environment with $100$ dual-path clusters uniformly distributed over all angles. The $\kappa$-factor is $0$\,dB, so the LOS path is barely stronger than the scattered paths. All other parameters are unchanged compared to the last figure. Fig.~\ref{figure_simulationNLOScase} shows the average sum rate for different user locations for the subcarriers $1$ to $300$.

We notice that the array geometry plays a much smaller role under rich scattering.
Movable antennas provide significant benefits when there are few subcarriers, but these gains vanish rapidly as $S$ increases. The reason is that the channel becomes highly frequency selective, so different subcarriers require different antenna positions to mitigate interference. Hence, movable antennas only perform closely to the interference-free upper bound when $S$ is small.
For practical subcarrier numbers, movable antennas only provide 2.4$\%$ larger rates than the sparse ULA. 
This suggests that the advantages of movable antennas appear primarily in LOS-dominant wideband scenarios, where the channel is similar on all subcarriers.

\subsection{Impact of Hardware Impairments}

The asymptotic performance limit in \eqref{eq:upper-limit} is $113$\,bit/s/Hz in the previous simulations with $\EVM=0.02$. Since the achieved sum rates were substantially lower, hardware impairments were not the main performance-limiting factor.

\begin{figure} 
        \centering 
        {\includegraphics[width=1\textwidth]{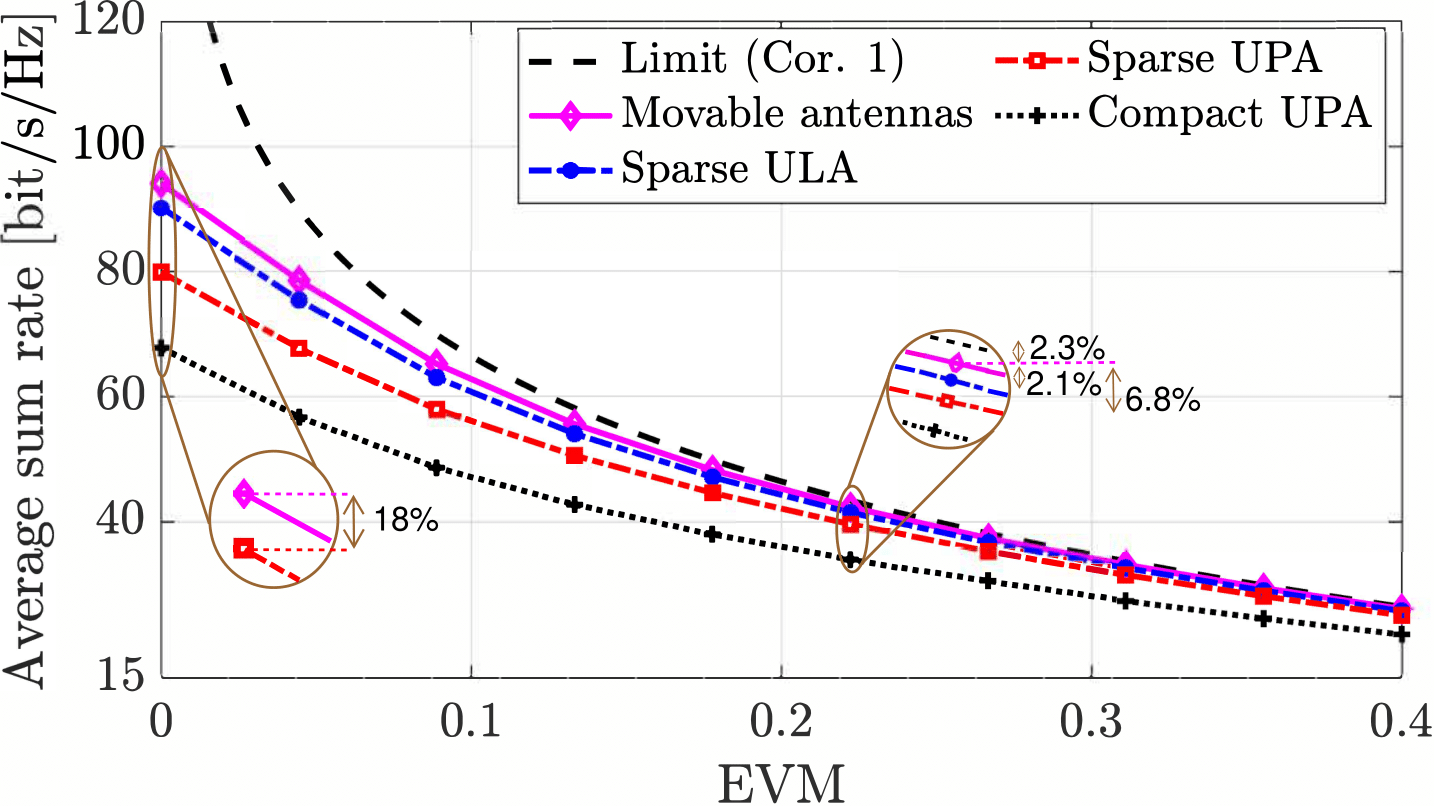}}
        \caption{Average sum rate versus EVM with $\rho=316$ mW.}
\label{figure_simulationEVM}  \vspace{-2mm}
\end{figure}

We will now revisit the setup from Fig.~\ref{figure_simulationLOScase} with $S=200$ subcarriers and a constant transmit power of $316$ mW. Fig.~\ref{figure_simulationEVM} shows how the average sum rate depends on the EVM. As the EVM increases, the sum rate for all schemes decreases and converges to the theoretical limit established in Corollary~\ref{cor1}. This confirms that hardware impairments degrade performance across all array configurations without benefiting movable antennas over fixed ones (which was not observed in \cite{yao2024rethinkinghardwareimpairmentsmultiuser}).

\begin{figure} 
        \centering 
        {\includegraphics[width=1\textwidth]{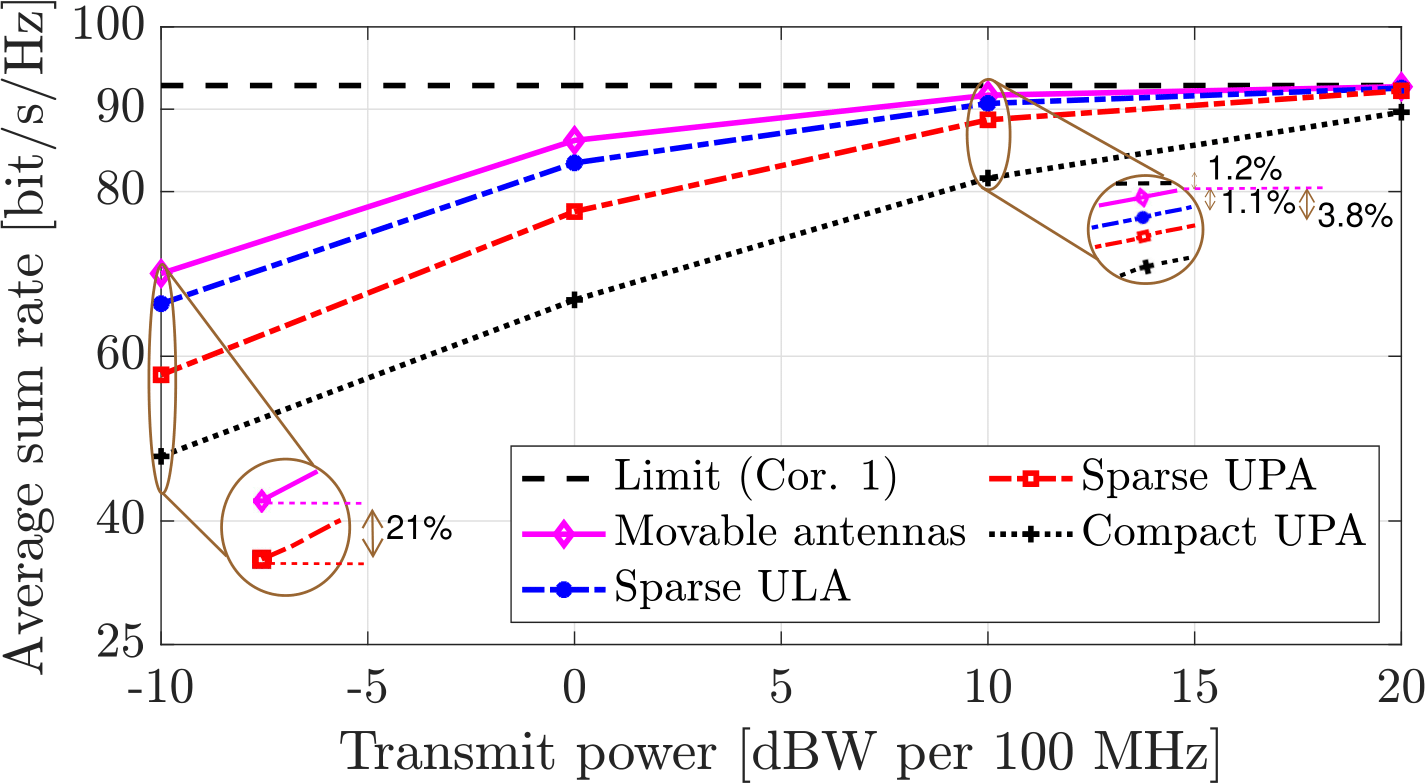}} \vspace{-6mm}
        \caption{Average sum rate as a function of the transmit power with $S=200$ subcarriers.} 
\label{figure_simulationSNR}
\end{figure}

Next, we examine the same setup with a constant $\EVM=0.04$, which limits the maximum achievable sum rate to $93$\,bit/s/Hz.  Fig.~\ref{figure_simulationSNR} shows the average sum rate as a function of the transmit power. As the transmit power increases, the average sum rate rises and eventually converges to the theoretical limit from Corollary~\ref{cor1}. While this limit remains the same for all array configurations, a well-designed array enables faster convergence.
However, as the SNR increases, the performance gap between different array configurations diminishes, indicating that the advantage of movable antennas becomes negligible at high SNRs. This suggests that movable antennas are most beneficial in interference-limited scenarios, whereas their performance gain over fixed antennas is minimal in hardware impairment-dominated conditions.

\vspace{-1.5mm}
\section{Conclusion}
\label{sec:conclusions}

This work quantifies the benefit of using movable antennas in a realistic wideband multi-user MIMO system. We derive a new OFDM system model with transceiver hardware impairments at the user side and a multipath propagation channel.

Our study concludes that the improvement in the average data rate with optimized movable antennas over same-sized sparse arrays decreases from around {\color{black}$22\%$} in LOS-dominant channels to no more than {\color{black}$5.3\%$} in non-LOS conditions.
Moreover, the relative gain from using movable antennas degrades with the number of subcarriers: from {\color{black}$14\%$} in a narrowband system with $\leq 20$ subcarriers down to {\color{black}$4.8\%$} in a wideband setup with $\geq 100$ subcarriers (in a non-LOS scenario). This happens because different subcarriers prefer different antenna positions.
In scenarios where movable antennas outperform a sparse ULA/UPA, one could potentially use a fixed irregular array optimized for the propagation scenario (as proposed in \cite{PIA2025}) to achieve similar rates without antenna movements.

We also notice that the benefit of using movable antennas is only significant when the inter-user interference dominates over the hardware distortion. 
 By contrast, with high transmit power levels, the presence of non-negligible hardware impairments starts diminishing any gain of movable antennas versus the performance achievable with any fixed antenna setup.

%\balance
\bibliographystyle{IEEEtran}
\bibliography{IEEEabrv,mybib}

\end{document}